# From Coffee Rings to Self-Driven Assembly: Active Matter Enabled Design of Drying Droplets

*Meneka Banik and Ranjini Bandyopadhyay**

Soft Condensed Matter Group, Raman Research Institute, C. V. Raman Avenue, Sadashivanagar, Bangalore 560080, India.

**ABSTRACT.** Evaporating colloidal droplets have long been used as model systems to understand capillarity, interfacial transport, and particle assembly, most prominently through the coffee ring effect. In classical descriptions, suspended particles are treated as passive tracers carried by evaporation-driven capillary flow, with additional influence from Marangoni stresses, wettability, and contact line pinning. More recent studies, however, show that this picture changes significantly when the particles themselves are active. Systems containing motile microorganisms, chemically active colloids, or externally driven particles can continuously inject energy or generate gradients within the droplet, leading to self-driven flows, modified interfacial stresses, and dynamic contact line behavior. In this Perspective, we bring together these developments, identify the key mechanisms governing active droplets, highlight the role of bubble-mediated flows, and outline strategies for controlled deposition and functional interface design.

## 1. INTRODUCTION.

### 1.1 Coffee ring Effect and Classical Framework:

Evaporation of particle-laden droplets has emerged as a versatile platform for probing interfacial transport, capillarity, and pattern formation. In the simplest case of passive suspensions, drying is governed primarily by evaporation-driven flows and boundary conditions at the contact line. A defining feature of such systems is the formation of the coffee ring deposit, in which particles accumulate at the droplet perimeter.[1,2] As illustrated in Figure 1(a), evaporation is spatially nonuniform and is maximized near the contact line. To replenish this enhanced evaporative loss, liquid is drawn radially outward from the droplet interior, establishing a capillary flow that transports suspended particles toward the edge.[2,3] When the contact line remains pinned, this outward flux persists throughout the drying process, leading to a continuous accumulation of particles at the periphery. Therefore, a dense, annular ring is formed at the contact line, accompanied by a particle-depleted central region. The sharpness and thickness of the ring are controlled by the strength and duration of the outward capillary flux, as well as by contact line pinning and evaporation rate.[1,2]

Deviations from this classical behavior arise when additional interfacial stresses are present. In particular, surface tension driven Marangoni flows can generate recirculating motion that opposes or redistributes the outward capillary transport, leading to more uniform or centrally accumulated deposits.[4,5] Taken together, passive droplets represent a boundary-driven transport regime in which particle trajectories are largely dictated by evaporation-induced capillary flow, and the resulting deposition morphology serves as a direct signature of this mechanism.

### 1.2 Breakdown of the Passive Paradigm - Role of Biological Matter:

Despite its success, the classical framework for evaporating colloidal droplets relies on the key assumption that suspended particles behave as passive tracers and do not actively influence the surrounding flow. This picture breaks down in the presence of active matter, where energy is continuously injected at the level of individual particles or organisms, introducing new transport pathways that reshape drying dynamics. As illustrated in Figure 1(b), droplets containing motile microorganisms such as bacteria exhibit a competition between evaporation-driven capillary flow and activity-induced transport.[6] In addition to the outward radial flux, self-propulsion introduces particle-scale advection (indicated by green arrows) that redistributes material throughout the droplet. As a result, edge accumulation is weakened, reorganized, or even replaced by more uniform or evolving deposition patterns.[6-8] In particular, motility near the contact line has been shown to influence how the deposit forms and evolves over time.[7] The final morphology reflects the relative strength of motility and evaporation-driven transport, as well as the spatial organization of the motile particles during drying.

Beyond motion, biological activity also brings chemical and interfacial effects into play. Many microorganisms secrete biosurfactants or extracellular polymeric substances (EPS), which can dynamically alter local surface tension.[6,9] These secretions create spatiotemporal gradients that drive Marangoni flows, often opposing or redirecting the outward capillary transport. Unlike externally imposed gradients, these flows emerge from the system itself and evolve with microbial activity and organization.[11] At higher microbial concentrations, collective behavior becomes important; dense suspensions can exhibit swarming, clustering, or even active turbulence, leading to enhanced mixing and large-scale recirculation.[11-13] Such collective dynamics disrupt the coherence of evaporation-driven transport and can produce irregular, patchy, or centrally concentrated deposits that differ strongly from classical coffee ring patterns.

In summary, biological droplets represent an internally driven transport regime in which particle trajectories and deposition patterns are governed by the interplay between capillary flow, active motion, and dynamically evolving interfacial stresses. The resulting morphologies - broadened rings, heterogeneous deposits, and partial suppression of edge accumulation - serve as clear signatures of a breakdown of the passive paradigm.

**1.3 Active Droplets:**

A widely studied class of active systems arises in evaporating droplets containing Janus particles; colloids with anisotropic surface properties that enable directed motion under external or self-generated gradients.[14-16] A broad range of propulsion mechanisms has been demonstrated, including light-activated thermophoretic motion, electrically-driven induced-charge electrophoresis, and magnetically-actuated propulsion.[17-19] In these systems, asymmetric stresses at the particle surface translate into self-propulsion and fluid disturbance at the microscale.[20,21] When present in evaporating droplets, active particles fundamentally alter the classical balance between capillary flow and interfacial stresses.[22,23] At the particle scale, self-propulsion may introduce active advection that can redistribute particles independently of evaporation-driven flow.[22-24] At higher particle concentrations, collective effects can give rise to coherent structures, including vortices, jets, or enhanced mixing, which perturb the radially outward transport characteristic of passive systems.[12,13] Beyond propulsion, many active Janus systems also modify interfacial properties, either directly or indirectly.[25-27]

One well-defined model system for investigating active particle dynamics is provided by catalytically active colloids. In chemically active droplets containing hydrogen peroxide ($H_2O_2$), catalytic decomposition at the particle surface (polystyrene-platinum, PS-Pt) produces spatially nonuniform fuel concentrations through asymmetric surface reactions.[23] At the droplet scale, consumption of $H_2O_2$ near regions of high catalytic activity establishes surface tension gradients along the liquid-air interface, driving Marangoni flows directed from low to high surface tension regions (Figure 1(c)).[28] These interfacial stresses generate recirculating flow structures that can oppose or redirect the outward capillary flux, promoting inward transport and enhanced mixing within the droplet.[23,28] Although catalytic decomposition of $H_2O_2$ produces oxygen gas, at low concentrations, visible bubbles are not present in the system. The various flows at this stage are indicated in Figure 1(c). However, at higher concentrations, oxygen nucleates as microbubbles within the droplet, and the flow transforms to a bubble-mediated, strongly catalytically active regime (Figure 1(d)).[29-31] As bubbles nucleate, grow, and migrate, their expansion displaces fluid, while collapse events induce rapid, short-lived recirculation and pressure fluctuations.[29-33] Unlike coherent Marangoni-driven flows, bubble-induced flows are inherently intermittent and spatially heterogeneous, continuously disrupting the internal flow field.[29,30]

The interplay between these mechanisms gives rise to a characteristic set of morphological outcomes. Instead of the well-defined peripheral ring observed in passive droplets (Figure 1(a), bottom panel) or the broadened deposits seen in biological systems (Figure 1(b), bottom panel), chemically active droplets produce highly dynamic and reconfigurable deposition patterns (Figures 1(c) and 1(d), bottom panels).[29,30] Deposits may exhibit asymmetric distributions, radial protrusions, patchy or discontinuous rings, and particle accumulation within the droplet interior.[29,30] Figure 1 shows the transition from boundary-driven to internally driven transport in evaporating droplets. While phoretic and Marangoni effects can generate sustained or slowly varying flows, bubble-mediated processes introduce strong temporal intermittency, leading to fluctuations in contact line dynamics, local depinning events, and temporal redistribution of particles.[23,29,30] Schematic illustrations are drawn by the authors and adapted conceptually from prior studies on passive, biological, and active droplet systems.[1,6,29]

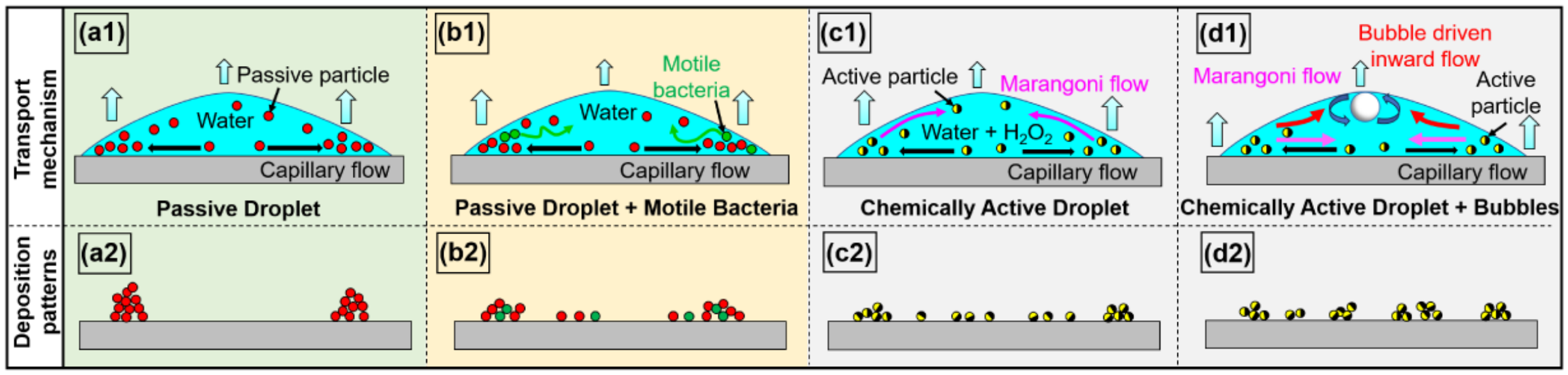

**Figure 1.** Comparison of transport mechanisms (top-panel) and resulting deposition patterns (bottom-panel) in evaporating aqueous droplets. (a) Passive droplets: evaporation-driven capillary flow (black arrows) transports particles to the pinned contact line, producing ring-like deposits with edge accumulation and a depleted interior. (b) Droplets with motile microorganisms: active motion redistributes particles against capillary flow (green arrows), leading to broadened or fragmented rings and heterogeneous deposits. (c) Chemically active droplets (PS-Pt Janus particles in lower $H_2O_2$ concentration): Marangoni stresses (pink arrows) and (d) Bubble-mediated flows (red and blue arrows) disrupt radial transport, producing asymmetric, discontinuous, or interior-rich deposits.

## 2. PERSPECTIVE.

### 2.1 Why a New Framework Is Needed:

In active systems, particles capable of self-propulsion, chemical reactions, or biological motion alter both internal flow fields and interfacial dynamics.[11,22,23] Transport in active droplets is therefore a continuous and coupled interplay between solvent evaporation, activity-induced motion, and interfacial response. This coupling gives rise to flow structures, contact line dynamics, and deposition patterns that deviate qualitatively from classical predictions.[11,22,23,29,34]

## 2.2 Key Parameters Controlling Deposit Morphologies in Active Droplets:

### 2.2.1 Evaporation rate:

Under rapid evaporation, outward capillary flow dominates for most of the drying process, favoring peripheral particle accumulation and ring-like deposits even in weakly active systems.[1-3,23,35] As evaporation slows, internally generated flows persist over longer timescales and increasingly redistribute particles away from the contact line.[6,7,11] In biological droplets, prolonged drying can enhance the influence of motility and collective rearrangements, producing broader and more spatially heterogeneous deposits.[6,7,12,13] In chemically active systems, slower evaporation allows concentration gradients and interfacial stresses to develop more fully, increasing the likelihood of inward redistribution and partial suppression of the coffee ring morphology.[4,23,28-30] In bubble-generating droplets, reduced evaporation rates are expected to prolong bubble lifetime and sustain repeated vortex-like recirculation, promoting patchy, asymmetric, or interior-rich deposits.[29,30] For externally driven active colloids, the evaporation rate may determine whether the imposed activity remains a perturbation to capillary flow or becomes the dominant transport mechanism. For example, in light-activated particles, slow evaporation combined with sustained photothermal forces may progressively shift deposition from edge-dominated rings toward centrally concentrated structures due to persistent inward recirculation.[36-39]

### 2.2.2 Activity strength:

Activity strength governs the degree to which internally generated transport perturbs evaporation-driven flow. In biological droplets, increasing motility can progressively disrupt coherent outward transport, leading to enhanced mixing and increasingly heterogeneous deposits.[6,7,11,13] At sufficiently high reaction rates in chemically active Janus systems, bubble nucleation introduces intermittent hydrodynamic forces that can dominate internal transport locally.[23,29,30,32,33] This framework may be extended to a broader range of active colloids in which the activity strength can be dynamically controlled through external fields.[15,17] In light-activated systems, increasing illumination intensity is expected to strengthen local thermophoretic or photothermal flows, progressively suppressing capillary transport, which may transform the deposit morphology from ring-like to centrally accumulated structures.[17,35,36] In magnetically active particles, stronger magnetic driving or field-induced alignment may generate directional collective motion and anisotropic circulation, leading to elongated or orientation-dependent deposition patterns rather than radially symmetric deposits.[40,41] Electrically driven active colloids are expected to produce field-dependent electrohydrodynamic vortices, where increasing activity strength could destabilize steady outward transport and generate periodic or oscillatory redistribution of particles near the contact line.[18,42] Unlike catalytic bubble-generating systems, where the intermittency arises from bubble nucleation and collapse, externally driven active systems may allow precise tuning of deposition morphology through dynamic on-off control of the applied stimulus.[17-19]

#### 2.2.3 Wettability:

Wettability controls deposit morphology by regulating contact line stability and the coupling between internally generated flows and the substrate interface.[1-5,29,30] Hydrophilic surfaces favor stronger contact line pinning of aqueous droplets and more axisymmetric deposition, whereas hydrophobic substrates promote higher contact angles and longer-lived internal circulation. In biological droplets, wettability can influence microbial adhesion and the localization of active clusters near the contact line.[6,9,10] In chemically active systems, hydrophobic interfaces may stabilize interfacial concentration gradients and prolong Marangoni-driven redistribution, whereas hydrophilic surfaces can suppress sustained recirculation.[4,23,28-30] Wettability becomes particularly important in bubble-generating systems because bubble-pinning and residence time depend strongly on surface energy.[29,30] Hydrophobic substrates stabilize bubbles and sustain localized vortex-like flows, producing fragmented rings, asymmetric deposits, and enhanced interior accumulation. In contrast, hydrophilic substrates favor rapid bubble detachment or dissolution, leading to weaker perturbations and more uniform deposition patterns. Because active colloids are highly sensitive to hydrodynamic and chemical boundary conditions, wettability may additionally determine how externally imposed flows couple to near-wall transport.[11,22,25,26,43]

### 2.2.4 Surface patterning:

Surface patterning provides spatial control over deposition morphology by locally modifying pinning forces, confinement, and activity-induced flow organization.[30] In passive droplets, patterned substrates alter ring symmetry and particle packing through geometric constraints on contact line motion.[1-5,35] In active systems, these effects become more pronounced because topography also interacts with internally generated transport.[22,23,29,30] In biological droplets, patterned surfaces may be expected to guide collective migration or localize active assemblies, leading to preferential accumulation along topographic features.[11-13,36] In chemically active systems, surface structures distort interfacial stress distributions, producing anisotropic particle redistribution.[23,28-30] Most significantly, in bubble-generating droplets, micro- and nanoscale roughness can localize bubble nucleation and suppress bubble mobility, thereby modifying the strength and spatial extent of vortex-driven transport.[29-33] Depending on the pattern geometry, this may either stabilize deposition symmetry or promote strongly heterogeneous structures with localized aggregation zones. For externally driven active colloids, patterned interfaces may guide particle motion and localize internally generated flows during evaporation.[36-42] As a result, substrate topography could be used to direct particle accumulation into specific regions, producing anisotropic or spatially programmed deposition patterns. Light-responsive particles may undergo thermally directed redistribution over patterned substrates due to locally generated temperature gradients, although the direction of migration depends on the particle-solvent interactions and thermophoretic response.[17,37-39] In contrast, magnetic or electrically driven colloids could exhibit field-guided assembly along structured interfaces.[40-42,44,45]

A generalized schematic summary of how evaporation rate, activity strength, wettability, and surface patterning influence deposition morphologies across biological, chemically active, and bubble-mediated systems is shown in Figure 2, based on trends reported in prior experimental and theoretical studies.[7,12,13,29,30]

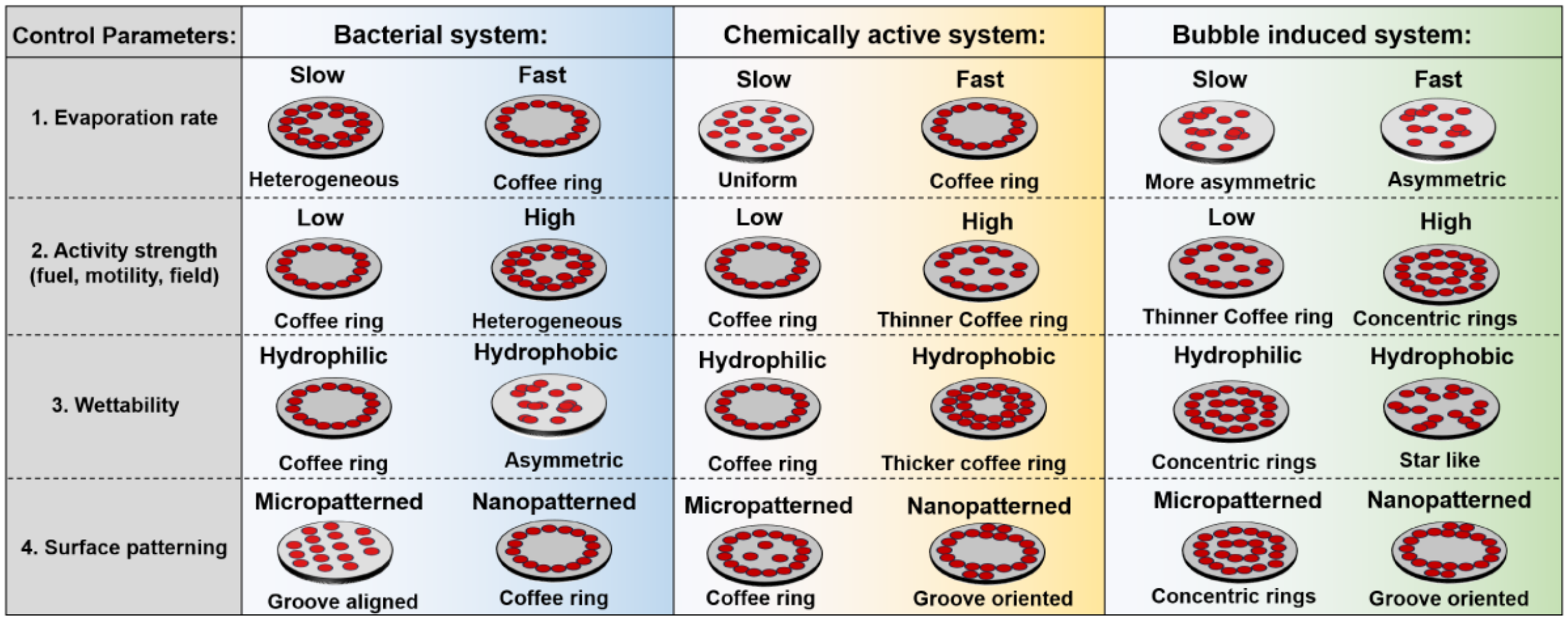

**Figure 2.** Schematic illustration of how key control parameters influence deposition morphologies in active aqueous droplets. Variations in evaporation rate, activity strength, substrate wettability, and surface patterning alter the balance between outward particle transport and inward recirculation, leading to distinct morphologies, including coffee ring, uniform, asymmetric, concentric, and groove-oriented deposits. Bubble-mediated systems exhibit the strongest morphological diversity due to transient vortex-like recirculation arising from bubble nucleation, growth, pinning, and collapse.

**2.3 From Local Dynamics to Emergent Deposit Morphologies:**

Recent studies suggest that deposition morphologies in active droplets are governed not only by global flow fields but also by localized dynamic events occurring over short spatial and temporal scales.[22,23,29,30] In contrast to passive droplets, where deposition correlates primarily with evaporation-driven capillary transport,[1-5] active systems are sensitive to transient recirculation, local particle orientation, and interfacial fluctuations.[22,23,25,26,29,30] Consequently, subtle local perturbations can produce large changes in the final deposit morphology.

In chemically active droplets (at high $H_2O_2$ concentrations), we recently showed that bubble-pinning and droplet-pinning strongly influence deposition pathways.[29,30] Pinned bubbles generate sustained localized recirculation that redistributes particles around the pinned site, producing corrugated rings, asymmetric deposits, and heterogeneous particle clusters.[29] In contrast, rapidly detaching or short-lived bubbles generate only transient perturbations and produce comparatively weaker morphological distortions. On micropatterned substrates, long-lived bubbles sustain droplet scale vortices, whereas on nanopatterned substrates, bubbles remain confined within nanoscale grooves and fail to develop large-scale recirculation.[30] We further showed that droplet pinning time decreases systematically with increasing catalytic activity on nanopatterned substrates due to repeated interfacial perturbations, while prolonged pinning on micropatterned substrates sustains bulk redistribution. Together, these observations identify bubble lifetime and droplet pinning time as key dynamic metrics governing morphology selection in active evaporating droplets.

Analogous localized metrics are also likely to emerge in chemically active (low $H_2O_2$ concentration) and biological systems. In chemically active Janus systems without bubble formation, quantities such as local fuel depletion rate, persistence length of active trajectories, transient Marangoni recirculation strength, and contact line residence time may similarly govern whether deposits remain ring-like, uniform, or spatially heterogeneous.[22,23] In bacterial droplets, surfactant production, wettability modification, and osmotic volume growth can dynamically alter contact line pinning and reorganize nearby transport pathways during spreading and evaporation.[6-13,46] Such effects may continuously perturb local capillary balance and generate heterogeneous particle redistribution even under otherwise symmetric drying conditions.[6-8,11-13] Comparable local perturbations may also arise in externally driven active colloids, where localized thermophoretic or phoretic recirculation can redirect nearby particle trajectories and suppress edge accumulation during evaporation.[17,37-39] Together, these observations clearly indicate that morphology selection in active droplets may generally be governed by evolving local dynamic metrics, rather than by evaporation-driven transport alone.

Figure 3 summarizes how localized dynamic events reported in recent active droplet studies correlate with final deposition morphologies.[22,23,29-31,36-39,46]

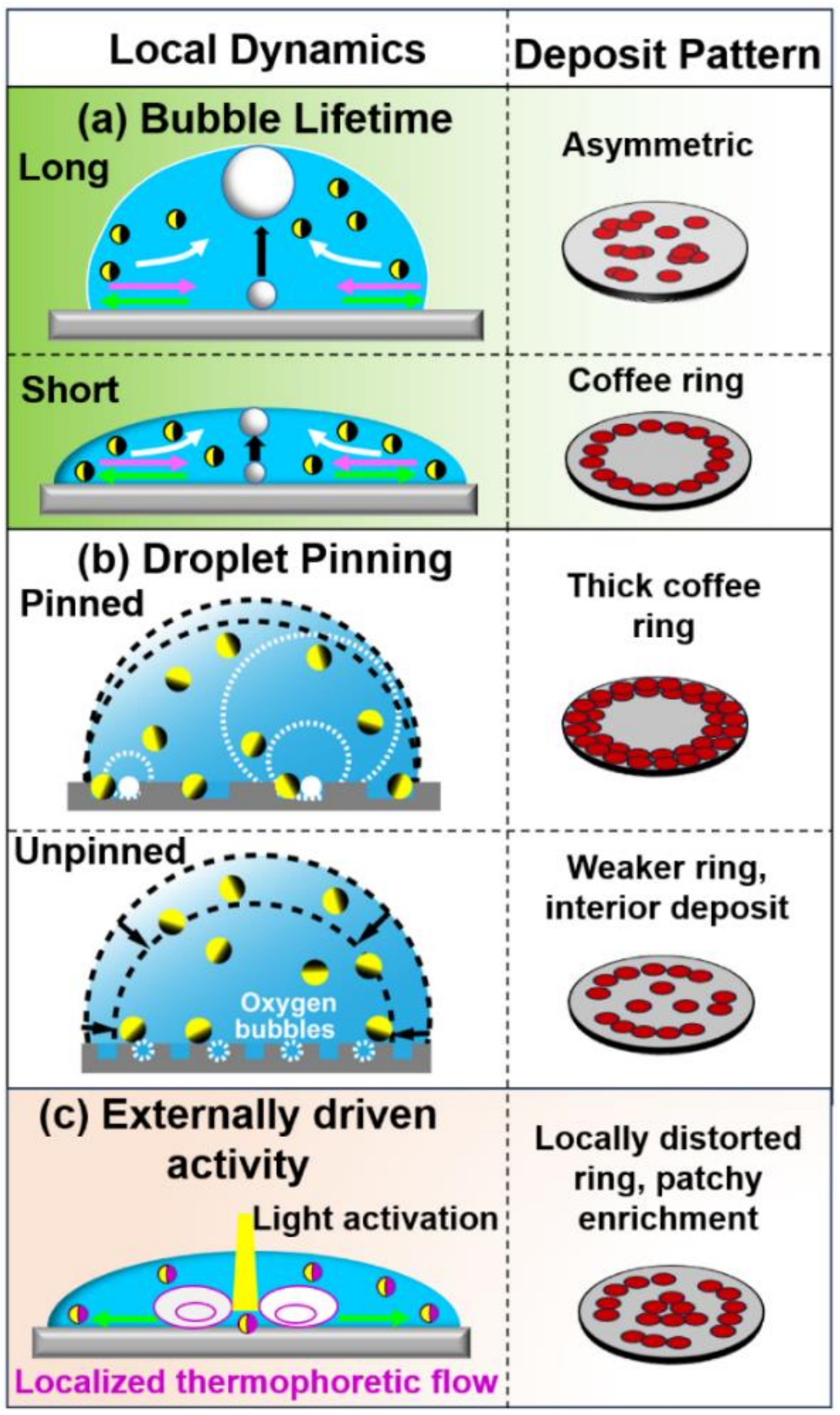

**Figure 3.** Correlation between localized dynamic events and final deposition morphologies in active droplets. (a) Bubble lifetime, (b) droplet pinning, and (c) externally driven local activity strongly influence particle redistribution during evaporation.

## 3. MOVING FORWARD.

### 3.1 Modeling Challenges:

Existing models for confined active films and bacterial suspensions typically describe the flow using Stokes hydrodynamics coupled with a continuous active force term,[47-49]

$$-\nabla p + \mu \nabla^2 u + f_{act} = 0$$

where $p$ is the hydrodynamic pressure, $\mu$ is the dynamic viscosity of the fluid, $u$ is the fluid velocity field and $f_{act}$ represents internally generated stresses arising from motility, self-propulsion, or collective activity.[47-49] Physically, $f_{act}$ acts as a distributed momentum source that continuously injects mechanical stresses into the fluid, thereby destabilizing quiescent flow fields and generating spontaneous vortices, active recirculation, and confinement-induced hydrodynamic instabilities. Such models successfully capture active flow organization in films and channels with fixed geometries.

However, evaporating active droplets differ fundamentally because, within the thin-film approximation, the droplet height $h(r, t)$ (where $h(r, t)$ is the local droplet height at radial position r and time t) evolves according to mass conservation,[50]

$$\frac{\partial h(r,t)}{\partial t} + \nabla[h(r,t).u(r,t)] = -J(r,t)$$

where $u(r, t)$ is the depth-averaged in-plane velocity and $J(r, t)$ is the local evaporative flux.

Conventional evaporation models successfully capture coffee ring formation, contact line dynamics, and evaporation-driven particle redistribution in passive systems.[50] In active droplets, however, internally generated flows can reorganize particles over short spatial and temporal scales.[22,23,29-33,47-49] Models of chemically reactive thin-films have shown that reaction-induced concentration gradients can destabilize confined liquid films through solutal Marangoni stresses, generating localized height modulations, and nonuniform film morphologies even in the absence of evaporation.[51] Such studies highlight that active interfacial transport alone can reorganize thin-film hydrodynamics. Developing predictive models for active droplet evaporation requires theoretical frameworks that simultaneously incorporate evaporation, interfacial stresses, activity-induced transport, reaction kinetics, and localized hydrodynamic fluctuations within a single evolving system. A major challenge is to identify experimentally accessible quantities that correlate local dynamics with final deposition morphologies.

**3.2 Open Questions, Outlook and Opportunities:**

Recent studies, including those from our group on chemically active systems, suggest that quantities such as contact line residence time, local recirculation strength, persistence of active trajectories, particle orientation, and bubble lifetime in reactive systems may serve as physically meaningful metrics governing morphology selection.[22,23,29,30] Direct experimental measurement of these quantities remains difficult because the relevant dynamics evolve over short length and time scales and continuously reorganize during drying. Simultaneous mapping of internal flow fields, interfacial deformation, particle trajectories, and transient localized activity therefore represents an important experimental frontier in active droplet research.[22,23,29-33,47-49] Improved visualization and flow-mapping approaches capable of resolving these transient events will likely be essential for developing predictive frameworks that link active hydrodynamics to deposition morphologies.

Active evaporating droplets therefore provide a versatile platform for coupling nonequilibrium transport, interfacial hydrodynamics, and self-generated flows within microscale confined geometries.[52,53] The internally generated activity enables dynamic control over mixing, particle redistribution, and interfacial transport during drying.[11,22-30,47-49,52,53] Such behavior creates opportunities for applications in microscale reactors, adaptive coatings, directed self-assembly, sensing, and active material fabrication.[52-55] For example, sustained internal recirculation may enhance local mixing and mass transport in evaporating microreactors, while localized bubble-mediated vortices could enable spatially selective particle concentration and reaction enhancement.[54,55] Similarly, localized active flows and interfacial recirculation may concentrate particles or analytes near pinned bubbles or contact lines, enabling spatially selective enrichment for sensing applications.[54] Surfaces with varying wettability may further provide a route for programming active transport during evaporation. By selectively controlling local contact line pinning and evaporation flux, chemically patterned substrates could localize active recirculation within specific regions of the droplet footprint, thereby enabling patterned particle deposition and directed assembly.[56]

In materials fabrication, the ability to dynamically reorganize particle transport during drying could enable programmable deposition pathways, anisotropic assemblies, and hierarchical microstructures that are difficult to achieve using passive evaporation alone. Such control may be particularly relevant for evaporation-assisted additive manufacturing and printed microfabrication, where dynamically tunable deposit patterns could improve pattern fidelity, local material organization and multiscale structural control during solvent evaporation.[57] Externally driven active particles may further allow real-time tuning of morphology through light, magnetic, or electric fields during evaporation.[58] Schematic shown in Figure 4 is author-generated and conceptually inspired by prior studies on active matter, externally driven active colloids, programmable interfaces and active material systems.[36-45,52-58]

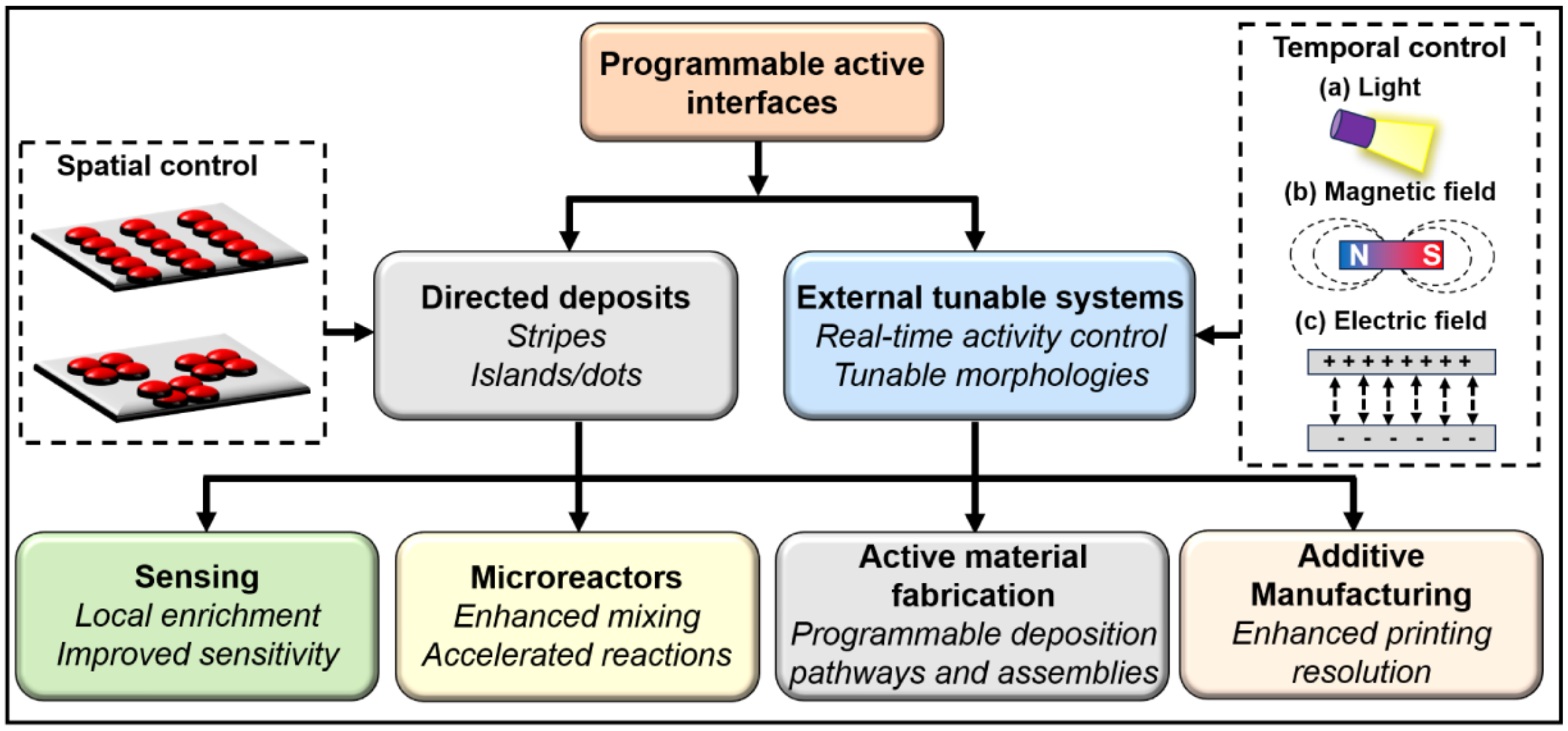

**Figure 4.** Outlook and opportunities for active evaporating droplets as programmable interfacial platforms.

From a fundamental perspective, active droplets also provide an experimentally accessible model system for studying nonequilibrium interfacial phenomena under evolving confinement. In particular, bubble-mediated systems introduce localized stochastic forces that can continuously reorganize global transport pathways during drying.[29-33,47-49]

## 4. Closing perspective.

Evaporating droplets have long been used to study interfacial transport and pattern formation, but the introduction of activity changes this picture in a fundamental way. Active droplets develop flows, which can significantly alter how particles move and where they eventually deposit. Across biological, phoretic, and chemically reactive systems, this internal activity competes with classical capillary and Marangoni effects, often leading to behavior that cannot be explained using traditional models alone. No single mechanism controls the outcome; rather, it is the balance between evaporation, interfacial forces, and internally generated activity that determines the observed flow patterns and final deposits. In this context, it is also useful to distinguish between systems where activity is sustained, such as in motile or phoretic particles, and those where it appears intermittently, as in bubble-generating systems. This difference alone can lead to very different transport pathways. Developing models that can capture these coupled processes remains difficult, especially given the range of length and time scales involved. Experimentally, resolving transient flows and quantifying features such as bubble lifetimes is also not straightforward. With further progress, it should be possible to move toward more predictive and controllable systems. By recognizing and harnessing these effects, droplet evaporation can be transformed from an apparently uncontrolled process into a tunable and versatile platform for pattern formation and interfacial engineering.